\documentstyle[sw20aip]{article}
 
\input{tcilatex}
\input tcilatex
\begin{document}
 
\title{A Class of Einstein-Maxwell Fields Generalizing the Equilibrium Solutions}
\author{Zolt{\'a}n Perj{\'e}s$^1$ \\
$1$ KFKI Research Institute for Particle and Nuclear Physics,\\
H--1525, Budapest 114, P.O.B.\ 49, Hungary}
\date{\today }
\maketitle
 
\begin{abstract}
The Einstein-Maxwell fields of rotating stationary sources are represented
by the SU(2,1) spinor potential $\Psi _A$ satisfying
\[
\nabla \cdot [\Theta ^{-1}(\Psi _A\nabla \Psi _B-\Psi _B\nabla \Psi
_A)]=-2\Theta ^{-2}\vec{C}\cdot (\Psi _A\nabla \Psi _B-\Psi _B\nabla \Psi
_A)
\]
where $\Theta =\Psi ^{\dagger }\cdot \Psi $ is the $SU(2,1)$ norm of $\Psi $%
. The Ernst potentials are expressed in terms of the spinor potential by $%
{\cal \ E}=\frac{\Psi _1-\Psi _2}{\Psi _1+\Psi _2}$, $\Phi =\frac{\Psi _3}{%
\Psi _1+\Psi _2}$ . The group invariant vector $\vec{C}=-2i\func{Im}\{\Psi
^{\dagger }\cdot \nabla \Psi \}$ is generated exclusively by the rotation of
the source, hence it is appropriate to refer to $\vec{C}$ as the {\em swirl}
of the field. Static fields have no swirl.
 
The fields with no swirl are a class generalizing the equilibrium ($\vert
e\vert =m$) class of Einstein-Maxwell fields. We obtain the integrability
conditions and a highly symmetrical set of field equations for this class,
as well as exact solutions and an open research problem.
\end{abstract}
 
 
\section{Introduction: {\bf Einstein-Maxwell fields of rotating stationary
sources}}
 
A stationary Einstein-Maxwell field is characterized by the two complex
Ernst potentials \cite{Ernst} ${\cal E}$ and $\Phi $ of the gravitational
and electromagnetic fields, respectively. For the investigation of the
internal symmetries of the system, it is more advantageous to use a pair of
new potentials $\xi $ and $q$, defined by the relations
\begin{equation}
{\cal E}=\frac{\xi -1}{\xi +1}\qquad \qquad \Phi =\frac q{\xi +1}.
\label{xiq}
\end{equation}
 
In terms of the $\xi $ and $q$ potentials, the Lagrangian of the
Einstein-Maxwell system takes the form
\begin{equation}
L=\theta ^{-2}[\partial _{\mu }\xi \partial ^{\mu }{\bar{\xi}}-\!\partial
_{\mu }q\partial ^{\mu }{\bar{q}}+(\xi \partial _{\mu }q-\!q\partial _{\mu
}\xi )({\bar{\xi}}\partial ^{\mu }{\bar{q}}-{\bar{q}}\partial ^{\mu }{\bar{
\xi}})]  \label{L}
\end{equation}
with
\[
\theta =\xi {\bar{\xi}}+q{\bar{q}}-1.
\]
 
Here the metric $g_{\mu \nu }$ is that of the Euclidean 3-space in any
suitable coordinate system. Hence the Euler-Lagrange equations follow:
\begin{eqnarray}
\theta \Delta \xi -2(\bar{\xi}\nabla \xi +\bar{q}\nabla q)\cdot \nabla \xi
&=&0  \label{feqs} \\
\theta \Delta q-2(\bar{\xi}\nabla \xi +\bar{q}\nabla q)\cdot \nabla q &=&0.
\nonumber
\end{eqnarray}
 
The global symmetries of this system form the $SU(2,1)$ group\cite
{Kinnersley}. In the next section, we shall briefly review the basic theory
of conserved currents and their relation to global symmetries. We then
compute the conserved currents of the Einstein-Maxwell system and find a
highly symmetrical form of the field equations using the currents. We
introduce the concept of the 'swirl vector' $\vec{C}$ which is a group
invariant. In the subsequent sections the fields satisfying the condition of
vanishing swirl are investigated.
 
\section{\bf Symmetries and Currents}
 
The action as a functional of the fields $\phi ^{i}$ and $\partial _{\mu
}\phi ^{i}$ is
 
\[
S=\int d^{n}x{\cal L}(\phi ^{i},\partial _{\mu }\phi ^{i}).\
\]
Associated with a symmetry transformation
 
\[
\phi ^{i}\to \phi ^{i}+\delta \phi ^{i}\ ,
\]
such that the Lagrangian density is invariant, $\delta {\cal L}=0\ ,$ and
\begin{equation}
\delta \phi ^{i}=\varepsilon _{k}^{i}\phi ^{k},
\end{equation}
there exists a conserved {\it \ }current
\begin{equation}
J^{\mu }=\frac{\partial {\cal L}}{\partial \left( \partial _{\mu }\phi
^{i}\right) }\varepsilon _{k}^{i}\phi ^{k}  \label{J}
\end{equation}
({\it Noether }current) satisfying
 
\[
\partial _\mu J^\mu =0\ .
\]
 
We now proceed with the application of this general theory to the
Einstein-Maxwell system. The Lagrangian $\left( \ref{L}\right) $ possesses
an $SU(2,1)$ global symmetry group \cite{Neugebauer}\cite{Kinnersley}. The
potentials $\alpha ,$ $\beta $ and $\gamma $ belonging to the fundamental
representation of this group are given by
\begin{equation}
\xi =\frac \alpha \beta \qquad \qquad q=\frac \gamma \beta .  \label{albega}
\end{equation}
The global invariance transformations are
\begin{equation}
\left(
\begin{array}{l}
\alpha \\
\beta \\
\gamma
\end{array}
\right) \to {\bf U}\left(
\begin{array}{l}
\alpha \\
\beta \\
\gamma
\end{array}
\right)
\end{equation}
where ${\bf U}\in SU(2,1)$ is a constant matrix.
 
The defining{\ representation of the global symmetry is given by the spinor
potential $\Psi _A=(\alpha ,\beta ,\gamma ),$ and its group adjoint $\Psi
^{\dagger A}=(\bar{\alpha},-\bar{\beta},\bar{\gamma})$}. Hence also the
Ernst potentials are written
 
\[
{\cal E}=\frac{\Psi _1-\Psi _2}{\Psi _1+\Psi _2}\qquad \qquad \Phi =\frac{%
\Psi _3}{\Psi _1+\Psi _2}.
\]
These spinors are determined up to an overall complex multiplying function;
the equivalence classes are given by the relation
\begin{equation}
{\Psi _A^{\prime }=\Omega \Psi _A}  \label{ray}
\end{equation}
where $\Omega $ is an arbitrary complex scalar.
 
The field equations $\left( \ref{feqs}\right) $ will take the $SU(2,1)$
invariant form
 
\begin{equation}
\nabla \cdot [\Theta ^{-1}(\Psi _{A}\nabla \Psi _{B}-\Psi _{B}\nabla \Psi
_{A})]=-2\Theta ^{-2}\vec{C}\cdot (\Psi _{A}\nabla \Psi _{B}-\Psi _{B}\nabla
\Psi _{A})  \label{feq}
\end{equation}
where $\Theta =\Psi ^{\dagger }\cdot \Psi $ is the $SU(2,1)$ norm of $\Psi $.
 
We now turn our attention to the group invariant vector
 
\[
\vec{C}=-2i\func{Im}\{\Psi ^{\dagger }\cdot \nabla \Psi \}.
\]
 
For static fields, all potentials are real, hence we have $\vec{C}=0$. We
thus infer that the vector $\vec{C}$ is a concomittant of the rotation of
the source, so that we will call $\vec{C}$ the{\it \ swirl} of the field.
The form of the vector $\vec{C}$ depends on the gauge (\ref{ray}). In this
paper, we consider Einstein-Maxwell fields for which it is possible to
choose a gauge in which the swirl vanishes:
 
\begin{equation}
\vec{C}=0.  \label{C0}
\end{equation}
In addition to static metrics, this condition characterizes the equilibrium (%
$|e|=m$) class of fields.
 
The currents $\left( \ref{J}\right) $ of the SU(2,1) symmetry can be
expressed by use of the swirl $\vec{C}$ as follows:
\begin{eqnarray}
J^{(1)} &=&\Theta ^{-2}(\alpha {\bar{\alpha}}+\beta {\bar{\beta}})\vec{C}%
+\Theta ^{-1}({\bar{\alpha}}\nabla \alpha -\alpha \nabla {\bar{\alpha}}+{\
\bar{\beta}}\nabla \beta -\beta \nabla {\bar{\beta}})  \nonumber
\label{eq:j1} \\
J^{(2)} &=&\Theta ^{-2}\alpha {\bar{\beta}}\vec{C}+\Theta ^{-1}({\bar{\beta}}%
\nabla \alpha -\alpha \nabla {\bar{\beta}})  \nonumber  \label{eq:j2} \\
J^{(3)} &=&\Theta ^{-2}\beta {\bar{\alpha}}\vec{C}+\Theta ^{-1}({\bar{\alpha}%
}\nabla \beta -\beta \nabla {\bar{\alpha}})  \nonumber  \label{eq:j3} \\
J^{(4)} &=&\Theta ^{-2}\alpha {\bar{\gamma}}\vec{C}+\Theta ^{-1}({\bar{\gamma%
}}\nabla \alpha -\alpha \nabla {\bar{\gamma}})  \nonumber  \label{eq:j4} \\
J^{(5)} &=&\Theta ^{-2}\gamma {\bar{\alpha}}\vec{C}+\Theta ^{-1}({\bar{\alpha%
}}\nabla \gamma -\gamma \nabla {\bar{\alpha}})  \label{eq:j5} \\
J^{(6)} &=&\Theta ^{-2}\beta {\bar{\gamma}}\vec{C}+\Theta ^{-1}({\bar{\gamma}%
}\nabla \beta -\beta \nabla {\bar{\gamma}})  \nonumber  \label{eq:j6} \\
J^{(7)} &=&\Theta ^{-2}\gamma {\bar{\beta}}\vec{C}+\Theta ^{-1}({\bar{\beta}}%
\nabla \gamma -\gamma \nabla {\bar{\beta}})  \nonumber  \label{eq:j7} \\
J^{(8)} &=&\Theta ^{-2}\gamma {\bar{\gamma}}\vec{C}+\Theta ^{-1}(\overline{%
\gamma }\nabla {\gamma }-{\gamma }\nabla \overline{\gamma }).  \nonumber
\label{eq:j8}
\end{eqnarray}
 
Here
 
\[
\Theta =\alpha {\bar{\alpha}}-\beta {\bar{\beta}}+\gamma {\bar{\gamma}}\label%
{eq:Thdef}
\]
and the swirl has the detailed form
 
\[
\vec{C}=\alpha \nabla {\bar{\alpha}}-\beta \nabla {\bar{\beta}}+\gamma
\nabla {\ \bar{\gamma}}-{\bar{\alpha}}\nabla \alpha +{\bar{\beta}}\nabla
\beta -{\bar{\gamma}}\nabla \gamma .\label{eq:cdef}
\]
 
From $J^{(1)}$ and $J^{(8)}$ we get
 
\begin{eqnarray*}
J^{(1a)} &=&\Theta ^{-2}\alpha {\bar{\alpha}}\vec{C}-\Theta ^{-1}(\alpha
\nabla {\bar{\alpha}}-{\bar{\alpha}}\nabla \alpha )  \label{eq:j1a} \\
J^{(1b)} &=&\Theta ^{-2}\beta {\bar{\beta}}\vec{C}-\Theta ^{-1}(\beta \nabla
{\bar{\beta}}-{\bar{\beta}}\nabla \beta ) .  \label{eq:j1b}
\end{eqnarray*}
 
For fields satisfying the condition of vanishing swirl, $\vec{C}=0,$ the
Einstein-Maxwell Eqs. (\ref{feq}) can be written in the simple form
 
\begin{equation}
\nabla \cdot [\Theta ^{-1}(\Psi _A\nabla \Psi _B-\Psi _B\nabla \Psi _A)]=0
\label{psipsi}
\end{equation}
From the equations (\ref{eq:j5}) of current conservation we get the
symmetrical set
 
\begin{equation}
\nabla \cdot [\Theta ^{-1}(\Psi ^{\dagger A}\nabla \Psi _B-\Psi _B\nabla
\Psi ^{\dagger A})]=0.  \label{cpsipsi}
\end{equation}
 
Equations (\ref{psipsi}) and (\ref{cpsipsi}) govern the class of
Einstein-Maxwell fields characterized by a vanishing swirl.
 
\section{\bf Formulation using the vectors $G$ and $H$:}
 
An Einstein-Maxwell system with one Killing vector may be fully
characterized by the complex 3-vectors \cite{Perjes}:
\[
{\bf G}=\frac{\nabla {\cal E}+2{\bar{\Phi}}\nabla \Phi }{2({\rm Re}{\cal E}+{%
\bar{\Phi}}\Phi )}\label{eq:Gdef},\qquad {\bf H}=\frac{\nabla \Phi }{({\rm Re%
}{\cal E}+{\bar{\Phi}}\Phi )^{1/2}}.
\]
In the notation referring to the metric of the three-space, the field
equations can be written
\begin{equation}
R_{\mu \nu }=-G_\mu {\bar{G}}_\nu -{\bar{G}}_\mu G_\nu +H_\mu {\bar{H}}_\nu +%
{\bar{H}}_\mu H_\nu  \label{eq:gheq0}
\end{equation}
 
\begin{eqnarray}
(\nabla -{\bf G})\cdot {\bf G} &=&{\bar{{\bf H}}}\cdot {\bf H}-{\ \bar{{\bf G%
}}}\cdot {\bf G}  \label{eq:gheq1} \\
(\nabla -{\bf G})\times {\bf G} &=&{\bar{{\bf H}}}\times {\bf H}-{\bar{{\bf G%
}}}\times {\bf G}  \label{eq:gheq2} \\
(\nabla -{\bf G})\cdot {\bf H} &=&\frac{1}{2}({\bf G}-{\bar{{\bf G}}})\cdot
{\bf H}  \label{eq:gheq3} \\
\nabla \times {\bf H} &=&-\frac{1}{2}({\bf G}+{\bar{{\bf G}}})\times {\bf H.}
\label{eq:gheq4}
\end{eqnarray}
 
The vectors ${\bf G}\ $and\ ${\bf H}$ can be expressed in terms of the
gradients of the complex potentials as follows,
\begin{eqnarray}
\Theta {\bf G} &=&\left( \overline{\xi }+1-q\overline{q}\right) \frac{\nabla
\xi }{\xi +1}+\overline{q}\nabla q \\
\Theta ^{1/2}{\bf H} &=&\left( \frac{\overline{\xi }+1}{\xi +1}\right)
^{1/2}\left( \nabla q-q\frac{\nabla \xi }{\xi +1}\right) .
\end{eqnarray}
Solving for the gradients, we have
\begin{eqnarray}
\nabla \xi &=&\Theta \frac{\xi +1}{\overline{\xi }+1}{\bf G-}\Theta ^{1/2}%
\overline{q}\left( \frac{\xi +1}{\overline{\xi }+1}\right) ^{3/2}{\bf H}
\label{grads} \\
\nabla q &=&\left( \overline{\xi }+1\right) ^{-1}q\Theta {\bf G+}\left(
\overline{\xi }+1-q\overline{q}\right) \Theta ^{1/2}q\left( \xi +1\right)
^{1/2}\left( \overline{\xi }+1\right) ^{-3/2}{\bf H.}  \nonumber
\end{eqnarray}
By use of the definitions $\left( \ref{albega}\right) $ and the vanishing of
the vector $\vec{C},$ we get the relation
\[
2i{\rm Im}\left( \frac{\xi \nabla {\bar{\xi}}+q\nabla {\bar{q}}}{\xi {\ \bar{%
\xi}}+q{\bar{q}}-1}\right) =\nabla \left( \ln \frac{\bar{\beta}}{\beta }%
\right) \label{eq:c0}.
\]
 
Inserting here the gradients $\left( \ref{grads}\right) ,$ we obtain the
integrability conditions of $\bar{\beta}/\beta $ in the form:
\begin{equation}
{\bf G\times \bar{G}=H\times \bar{H}.}  \label{intcond}
\end{equation}
{\ These constraints are apparently milder than the equilibrium ($R_{ij}=0$)
condition $Re({\bf G\otimes \bar{G}-H\otimes \bar{H})=0}$ characterizing the
Einstein-Maxwell fields with balanced electromagnetic and gravitational
forces. }
 
With the help of the integrability condition (\ref{intcond}), Eq. (\ref
{eq:gheq2}) takes the simple form
 
\begin{equation}
\nabla \times {\bf G}=0.  \label{eq:g}
\end{equation}
Hence there exists a complex potential $\psi $ such that
 
\[
{\bf G}=\nabla \ln \psi .\label{eq:g}
\]
 
Using this in Eq. (\ref{eq:gheq4}),
 
\[
\nabla \times {\bf H}=\frac 12\nabla (\ln \psi {\bar{\psi}})\times {\bf H.}
\]
Hence
 
\begin{equation}
{\bf H}=(\psi {\bar{\psi}})^{-1/2}\nabla \chi  \label{eq:h}
\end{equation}
for some complex function $\chi $. Eq. (\ref{eq:gheq0}) takes the form
 
\[
R_{\mu \nu }=-2\frac{\psi _{_{(,\mu }}{\bar{\psi}}_{,\nu )}-\chi _{_{(,\mu }}%
\overline{{\chi }}_{,\nu )}}{\psi {\bar{\psi}}}\label{eq:0}.
\]
 
The essential field equations can now be written down in terms of the
complex potentials $\psi $ and $\chi $ as follows,
 
\begin{eqnarray}
\nabla \cdot [\psi ^{-2}(\overline{{\psi }}\nabla \psi -\overline{{\chi }}%
\nabla \chi )] &=&0  \label{eq:1} \\
\nabla \cdot (\psi ^{-2}\nabla \chi ) &=&0  \label{eq:2} \\
\nabla \psi \times \nabla \overline{{\psi }}-\nabla \chi \times \nabla
\overline{\chi } &=&0.  \label{eq:3}
\end{eqnarray}
 
These are two complex and one real equations for the two complex functions $%
\chi $ and $\psi $. Despite this apparent overdetermination of the system,
at least two large classes of solutions exist. One of these consists of
stationary fields that are $SU(2,1)$ rotations of a static state. This is
due to the fact that the potentials are real for static fields and that the
vector $\vec{C}$ is $SU(2,1)$ invariant. The other subset of solutions of
these equations has the form $\chi =\psi .$ Field Eqs. (\ref{eq:1})-(\ref
{eq:3}) then reduce to the simple condition that $\psi ^{-1}$ is a harmonic
function. This is the equilibrium class characterized by a vanishing Ricci
tensor.
 
\section{Nonstatic fields}
 
An example of the $SU(2,1)$ global transformations generating a nonstatic
state from a given static electrovacuum is the Kramer-Neugebauer
transformation \cite{Kramer}, \cite{Esposito}. This has the form
\begin{equation}
\alpha ^{\prime }=(1-z\bar{z})\alpha ,\qquad \beta ^{\prime }=(1+z\bar{z}%
)\beta -2\overline{z}\gamma ,\qquad \gamma ^{\prime }=\left( \tfrac{%
\overline{z}}z\right) ^{1/2}\left[ (1+z\bar{z})\gamma -2z\beta \right]
\label{KN}
\end{equation}
where $z$ is a complex constant and $\left( \alpha ,\beta ,\gamma \right) $
is the real triplet of potentials for a static solution. Choosing $\left(
\alpha ,\beta ,\gamma \right) $ to be the potentials of an asymptotically
flat electrovacuum, the metric so obtained remains asymptotically flat. \
 
There is a wide range of static electrovacuum space-times which is suitable
to be chosen as the seed metric of global symmetry transformations. All
vacuum Weyl metrics and their electrovacuum counterparts, characterized by
the condition $q=const.,$ belong to these. A further simple case of an
explicit static electrovacuum is the {\sl Bonnorized} \cite{Bonnor} Kerr
metric in oblate spheroidal coordinates $\left( x,y\right) $:
\begin{equation}
\alpha =x^2+p^2-q^2y^2,\qquad \beta =2px,\qquad \gamma =2pqy
\end{equation}
where $p^2+q^2=1.$ This solution describes the gravitational and magnetic
field of a dipole source. An infinite sequence of static solutions is given
by the {\sl Bonnorized }Tomimatsu-Sato metrics\cite{aberdeen}. A wealth of
nonstationary solutions is readily available by the $SU(2,1)$ rotations of
these metrics. Among these, the rotating solutions obtained by the
Kramer-Neugebauer transformation (\ref{KN}) will possess controllable
asymptotic properties.
 
\section{Conclusions}
 
The Einstein-Maxwell fields with no swirl comprise a large number of
space-times, and among those ones with astrophysical significance. In
addition to the equilibrium ($|e|=m$) electrovacua, all static fields as
well as their stationary counterparts resulting from the global $SU(2,1)$
symmetry belong to this class. To be able to better assess the extent of
generality of this class, it would be clearly be of much help to find the
solution of the following open research problem:
 
{\it Find an example which is neither an equilibrium solution nor an }$%
SU(2,1)${\it \ rotation of a static field.}
 
Likewise, it would be of considerable interest to see if the Kinnersley\cite
{Kinnersley} group $K^{\prime }$ has a subgroup leaving the condition of
vanishing swirl invariant.
 
\section{Acknowledgments}
 
I thank M\'{a}ria S\"{u}veges and M\'{a}ty\'{a}s Vas\'{u}th for computing
the $SU(2,1)$ currents and for discussions. This work has been supported by
the OTKA grant T031724.

\end{document}